\documentclass{article}

    \usepackage[preprint]{neurips_2022}

\usepackage[table]{xcolor}
\usepackage[utf8]{inputenc} 
\usepackage[T1]{fontenc}    
\usepackage{hyperref}       
\usepackage{url}            
\usepackage{booktabs}       
\usepackage{amsfonts}       
\usepackage{nicefrac}       
\usepackage{microtype}      
\usepackage{xcolor}         
\usepackage[backend=biber,style=numeric,sorting=none,  bibstyle=numeric]{biblatex}
\usepackage{graphicx}
\usepackage{subcaption}
\usepackage{caption}
\usepackage{mathabx}
\usepackage{array}
\usepackage{makecell}

\title{A Survey on Evaluation Metrics for Synthetic Material Micro-Structure Images from Generative Models}

\author{%
  Devesh Shah \\
  Ford Motor Company\\
  \texttt{dshah22@ford.com} \\
      \And
   Anirudh Suresh \\
   Michigan State University \\
   \texttt{suresha2@msu.edu} \\
   \And
   Alemayehu Admasu \\
   Ford Motor Company \\
   \texttt{aadmasu@ford.com} \\
   \And
   \And
   Devesh Upadhyay \\
   Ford Motor Company \\
   \texttt{dupadhya@ford.com} \\
  \And
     Kalyanmoy Deb \\
   Michigan State University \\
   \texttt{kdeb@egr.msu.edu} \\
}

\begin{document}

\maketitle

\begin{abstract}
  
  The evaluation of synthetic micro-structure images is an emerging problem as machine learning and materials science research have evolved together. Typical state of the art methods in evaluating synthetic images from generative models have relied on the Fr\'echet Inception Distance. However, this and other similar methods, are limited in the materials domain due to both the unique features that characterize physically accurate micro-structures and limited dataset sizes. In this study we evaluate a variety of methods on scanning electron microscope (SEM) images of graphene-reinforced polyurethane foams. The primary objective of this paper is to report our findings with regards to the shortcomings of existing methods so as to encourage the machine learning community to consider enhancements in metrics for assessing quality of synthetic images in the material science domain.
  
\end{abstract}

\section{Introduction}

\label{section:intro}
The adoption of artificial intelligence techniques within the material science community has been critical in enabling high throughput research through rapid characterization and discovery of materials. In particular, various forms of generative models have been proposed to accelerate the design of new materials \cite{other_gans_for_matsci_nonImage, other_gans_for_matsci_nonImage2, ford_paper_currentworkplan, other_gans_for_matsci}. Nonetheless, the applications of traditional image based deep learning models in this domain are constrained, among other design choices, by the ability to evaluate trained models and synthetic image quality for physically accurate characteristics. 

Generative networks are unsupervised learning methods that aim to learn the regularities and patterns in input data such that the model can be used to generate new examples that could have plausibly been drawn from the original dataset \cite{goodfellow_gans}. These models are traditionally evaluated using the Fr\'echet Inception Distance (FID) score, which leverages the InceptionV3 model pretrained on the ImageNet dataset. Several studies have shown features from this model to be \textit{vision-relevant} \cite{fid_score_created}. By comparing the mean and variance of these features between synthetic and ground truth datasets, the FID score evaluates both the image quality and diversity of the synthetic images. Though plenty of works have highlighted various pitfalls of the FID score and suggested variations, it is still regarded as the standard for model evaluation. In this study, we explore the drawbacks of existing methods when working with datasets in the materials domain. In particular, we outline two main causes of concern: 

\begin{enumerate}
    \item Dataset Size: Materials scientists have worked with small \& noisy datasets with great success in uncovering process-structure-property relationships \cite{ford_paper_currentworkplan}. However, in small data regimes, the FID score promotes over-fitting, as the distance between distributions is minimized when matching the smaller variance of our ground truth data. Refer to Appendix \ref{appendix:dataset_limitations} for an experimental analysis.
    \item Feature Relevance: Unlike well-studied datasets (e.g. ImageNet), many of the features of a realistic micro-structure image cannot be recognized by the human eye, thus limiting the applications of a model pretrained on ImageNet as the base of the metric. Beyond image quality and diversity, in the material science domain, we aim to generate valid micro-structures which can be synthesized in an experimental setting.
\end{enumerate}

\section{Methods}

\label{section:methods}
Unlike traditional machine learning metrics, the evaluation of generative models and synthetic images have been an open subject of study for many years. Many works \cite{relatedWorks_pro_cons} have summarized the relationships between various metrics and outlined corresponding limitations. However, much of this work has been done on large datasets of simple images where population surveys can verify results. On the contrary, verification of synthetic images in the material science domain is limited to costly experimental trials \cite{gallagher2020predicting}. Subsequently, this limits the use of metrics using human evaluation such as those provided by Amazon Mechanical Turk \cite{amazon_turk} and HYPE \cite{hype_paper}. Studies in the medical domain \cite{medical_imaging_paper} have shown how standard metrics can give misleading results when applied to more complex images. 


Synthetic micro-structure images have significant potential to accelerate the materials discovery platform. SEM image data provides researchers with additional information about materials (e.g. connectivity of cells, cell sizes) that cannot be inferred by numerical data measurements. In particular, end-to-end inverse and forward structure-property models can allow us to replicate experimental testing of new materials computationally. Generative models, along with advances in methods such as attribute editing, allow us to extrapolate beyond our training dataset and build new variations of cell structures \cite{ford_paper_currentworkplan}. Corresponding auxiliary networks can then help classify the properties and process parameters of such images. This places the generation of synthetic data at the center of future materials discovery platforms. 

For our study, we use a dataset of SEM images of graphene-reinforced polyurethane foams (PUF) with seven unique additive amounts. The images were acquired under an SE2 secondary electron detector at fixed beam energy and working distance with 25x magnification (approx. 10.3 pix/micron). The brightness and contrast levels were held constant across all images and samples. 

There are three key properties we hope to evaluate in our synthetic micro-structure image datasets:

\begin{enumerate}
    \item Image Quality: Are the images of high fidelity? 
    \item Image Diversity: Are new images (outside the training data) being formed?
    \item Physically Accurate: Are the synthetic structures experimentally feasible/valid characteristics of materials (e.g. complete cell walls, inter-connected cell structures)?
\end{enumerate}

\begin{table}[!b]
\centering
    \vspace*{-4mm}
    \caption{Summary and sample images of the 4 datasets. State of the art methods \cite{stylegan2_implementation, diff_aug} were used.}
    \vspace*{1mm}
\begin{tabular}{ p{4cm} >{\centering\arraybackslash}m{2cm} 
>{\centering\arraybackslash}m{2cm}
>{\centering\arraybackslash}m{2cm}
>{\centering\arraybackslash}m{2cm}}
\hline
                       & Ground Truth Dataset & Dataset A           & Dataset B           & Dataset C               \\ \hline
Num. Images            & 2800                 & 1000                & 1000                & 1000                    \\
Data Generation Method & Experimental         & StyleGAN2 + DiffAug  & StyleGAN2 + DiffAug  & Variational Autoencoder \\
Sample Image           &      \includegraphics[width=0.15\textwidth]{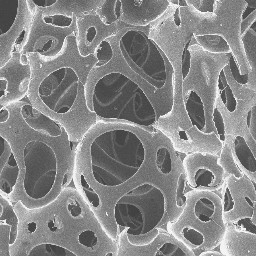}                &         
\includegraphics[width=0.15\textwidth]{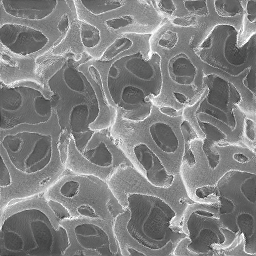}            & 
\includegraphics[width=0.15\textwidth]{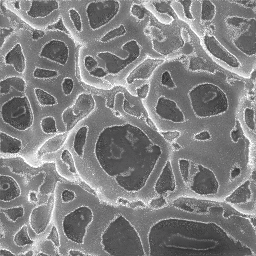} &
\includegraphics[width=0.15\textwidth]{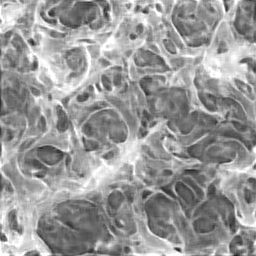} \\ \hline
\end{tabular}
    \label{tab:understanding_datasets}
\end{table}

Several generative models were trained using our ground truth dataset. In this study, we focus on the outputs from three particular models, summarized in Table \ref{tab:understanding_datasets} and detailed further in Appendix \ref{appendix:model_arch}. Resources were not spent to find ideal model parameters/architecture, as the focus was on comparing relative outputs of different generative models. In particular, outputs from Dataset B, though visually similar to Dataset A, are representative of mode-collapse.
Refer to Appendix \ref{appendix:image_analysis} for a detailed visualization and explanation of the image quality and diversity in each dataset.

Many alternatives of the FID score have been proposed and were considered during our study. These include the \textit{Inception Score (IS)} \cite{inception_score} and the \textit{Kernel Inception Distance (KID)} \cite{kid_score} to address concerns with the small training dataset. To differentiate between the image fidelity and diversity we also considered \textit{Precision \& Recall (P\&R)} \cite{precision_recall_orig, precision_recall} and \textit{Density \& Coverage} \cite{density_coverage}.

To mitigate the feature relevance concern with the FID score, various other feature extractors have been proposed, from randomly initialized weights \cite{density_coverage} to task-specific tuned models \cite{medical_imaging_paper}. We leveraged transfer learning methods to fine-tune an InceptionV3 model to perform a simple classification task of matching an input PUF image to one of the seven unique additive combinations. The model was trained for 10 epochs and achieved a >97\% accuracy on both the training (2000 images) and test (800 images) sets. Similar to the FID, the output of the final pooling layer was extracted as the 2048-dim feature vector for calculations. We define this as the \textit{FID Modified} score.

We additionally considered a wide array of image processing methods, from pixel-wise histograms to Image Quality Assessment (IQA) metrics. These included BRISQUE \cite{brisque_paper}, LPIPS \cite{lpips_paper, lpips_discussion}, Manipulative Precision Metric \cite{manipulative_precision_metric}, Optimal Transport \cite{optimal_transport_survey}, along with others. A key advantage of IQA methods is the ability to score individual images on image naturalness/distortion based on deviation from natural image distributions. However, experimental results showed these metrics were not suited for our application, with all our synthetic images scoring better than ground truth images.

\section{Results and Discussion}

Metric results from the FID score and alternatives are summarized in Table \ref{tab:metric_results}. We can infer Dataset A is consistently slightly higher ranked than Dataset B, and both Datasets A \& B are significantly better than Dataset C. Table \ref{tab:visual_inspection_of_data} provides a visual interpretation of each of the datasets for each of our key scoring properties we hope to evaluate. We observe the \textit{vision-relevant} features (scoring Dataset A the best) don't correlate well with our target of \textit{physically-accurate}. Similarly, Figure \ref{apdx:fig:image_diversity} highlights the severe over-fitting in Dataset B resulting in lack of any diversity - yet this dataset is still ranked similar to Dataset A. It is clear there are flaws in the existing methods. 

\label{section:results}
\begin{table}[!htbp]
    \vspace*{-2mm}
    \begin{minipage}{.65\linewidth}
      \centering
                  \captionsetup{width=0.9\linewidth}
              \caption{Standard metrics for each of the synthetic datasets relative to the ground truth. Average of 10 trials shown.}
              \vspace*{1mm}
\begin{tabular}{lccc}
\hline
\textbf{} & \multicolumn{3}{c}{Dataset} \\
\textbf{}       &  A &  B &  C \\ \hline
FID $(\downarrow)$          & \textbf{60 $\pm$ 0.4}        & 65 $\pm$ 0.3                 & 239 $\pm$ 1.7                \\
FID Modified $(\downarrow)$ & \textbf{97 $\pm$ 1.0}        & 208 $\pm$ 1.7                & 438 $\pm$ 2.8                \\
IS $(\uparrow)$           & \textbf{1.3 $\pm$ 0.0}       & 1.2 $\pm$ 0.0                & 1.2 $\pm$ 0.0                \\
KID $(\downarrow)$          & \textbf{0.1 $\pm$ 0.0}       & 0.1 $\pm$ 0.0                & 0.5 $\pm$ 0.0                \\
Precision $(\uparrow)$    & \textbf{0.5 $\pm$ 0.0}       & 0.3 $\pm$ 0.0                & 0.0 $\pm$ 0.0                \\
Recall $(\uparrow)$       & \textbf{0.0 $\pm$ 0.0}       & 0.0 $\pm$ 0.0                & 0.0 $\pm$ 0.0                \\
Density $(\uparrow)$      & \textbf{0.2 $\pm$ 0.0}       & 0.1 $\pm$ 0.0                & 0.0 $\pm$ 0.0                \\
Coverage $(\uparrow)$     & \textbf{0.04 $\pm$ 0.0}      & 0.02 $\pm$ 0.0               & 0.00 $\pm$ 0.0              \\ \hline
\end{tabular}
\vspace*{1mm}
              \label{tab:metric_results}
    \end{minipage}%
    \begin{minipage}{0.35\linewidth}
      \centering
      \captionsetup{width=0.9\linewidth}
            \caption{Visual inspection of each dataset. Complete analysis is shown in Appendix \ref{appendix:image_analysis}. Note: $\uparrow$ good, - neutral, $\downarrow$ bad}
            \vspace*{1mm}
\begin{tabular}{ p{1.5cm} >{\centering\arraybackslash}m{0.5cm} 
>{\centering\arraybackslash}m{0.5cm}
>{\centering\arraybackslash}m{0.5cm}}
\hline
\textbf{} & \multicolumn{3}{c}{Dataset} \\
                    &  A &  B &  C \\ \hline
\begin{tabular}[c]{@{}l@{}}Image \\ \hspace{0.2cm}Fidelity\end{tabular}      & \cellcolor{green!25}$\uparrow$         & \cellcolor{green!25}$\uparrow$         & \cellcolor{red!25}$\downarrow$         \\ 
\begin{tabular}[c]{@{}l@{}}Image \\ \hspace{0.2cm}Diversity\end{tabular}     & \cellcolor{yellow!25}-         & \cellcolor{red!25}$\downarrow$         & \cellcolor{green!25}$\uparrow$         \\ 
\begin{tabular}[c]{@{}l@{}}Physically \\ \hspace{0.2cm}Accurate\end{tabular} & \cellcolor{red!25}$\downarrow$         & \cellcolor{yellow!25}-         & \cellcolor{yellow!25}-        \\ \hline
\end{tabular}
\vspace*{1mm}
            \label{tab:visual_inspection_of_data}
    \end{minipage} 
\end{table}

\begin{figure}[!t]
\vspace*{-2mm}
    \centering
    \subfloat[\centering Ground Truth Dataset]{
        {\includegraphics[width=.49\linewidth]{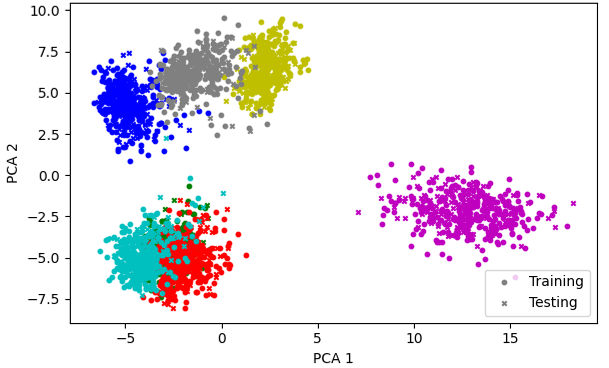} }
    }
    \enspace
    \subfloat[\centering Dataset A]{
        {\includegraphics[width=.49\linewidth]{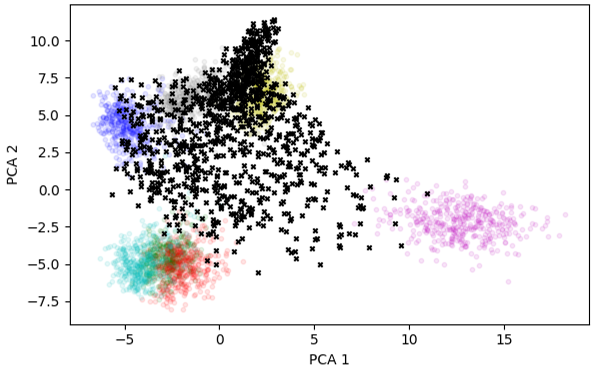} }
    }
    \\
    \vspace*{1mm}
    \subfloat[\centering Dataset B]{
        {\includegraphics[width=.49\linewidth]{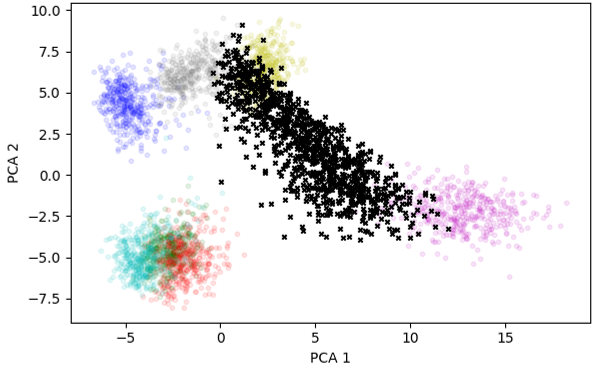} }
    }
        \enspace
    \subfloat[\centering Dataset C]{
        {\includegraphics[width=.49\linewidth]{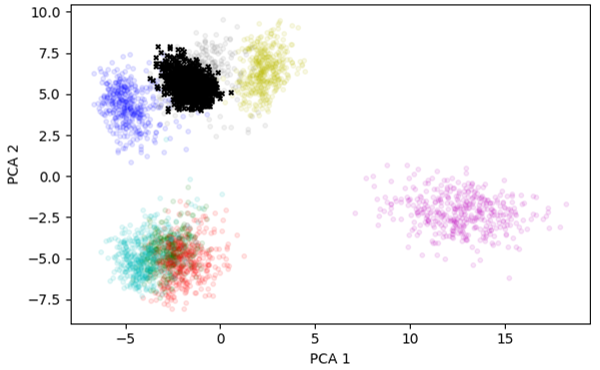} }
    }
    \\
    \vspace*{1mm}
        \subfloat{
        {\includegraphics[width=.6\linewidth]{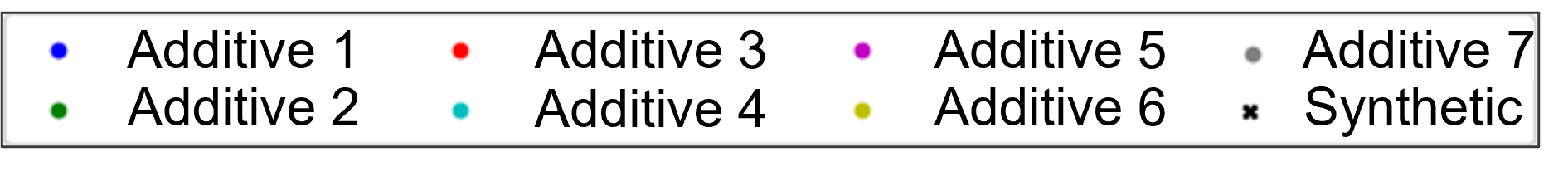} }
    }
    \vspace*{-2mm}
    \caption{2D PCA of feature vectors for each image using fine-tuned InceptionV3 model.}
    \label{fig:pca_results}
    \vspace*{-4mm}
\end{figure}

From Table \ref{tab:metric_results} we see our FID Modified score provides the greatest differentiation between datasets, with Dataset A being the best. However, there is an evident lack of value for the greater diversity seen in Dataset C. To further investigate, we extracted the outputs from the final pooling layer of the InceptionV3 model (fine-tuned on our materials dataset) to obtain 2048-dim feature vectors for each image. Using PCA, we were able to visualize these images in 2D, as seen in Figure \ref{fig:pca_results}. Note, for the ground truth dataset, we see the expected clusters for the seven additive types, as this is what the model was trained to distinguish. The black data points represent the synthetic images distributed on the same two principal components.

The results from Figure \ref{fig:pca_results} are quantified in Table \ref{tab:dist_results} as the euclidean distance between various sets of image feature vectors. For each of the four datasets, we calculated the \textit{same} and \textit{different} distances with respective to each additive type. The \textit{same} distance represents the distance between each point in Dataset X classified as Additive Y (using our fine-tuned InceptionV3 model) with every point in the Ground Truth Dataset of Additive Y. Conversely, the \textit{different} distance compares every point in Dataset X classified as Additive Y with every point in the Ground Truth Dataset not of Additive Y. Note these distances are calculated using the complete 2048-dim feature vectors. 

Similar to the 2D PCA in Figure \ref{fig:pca_results}, Table \ref{tab:dist_results} indicates there is a clear distinction between each category. For the Ground Truth Dataset, distances within the same category average 12.1 while distances to other category labels average 20.8. An ideal synthetic dataset should provide \textit{same} distance values between these endpoints - indicating both realistic images and greater diversity than the original categories. Results from Datasets A-C support this.

\begin{table}[htb]
\centering
    \vspace*{-2mm}
\caption{Distances between 2048-dim feature vectors of dataset images and corresponding additive types from training dataset. Any values with "-" indicate no synthetic images were generated with that category prediction for that dataset.}
    \vspace*{1mm}
\begin{tabular}{ccccccccc}
\hline
           & \multicolumn{2}{c}{Ground Truth Dataset} & \multicolumn{2}{c}{Dataset A} & \multicolumn{2}{c}{Dataset B} & \multicolumn{2}{c}{Dataset C} \\
           & Same             & Different             & Same        & Different       & Same        & Different       & Same        & Different       \\ \hline
Additive 1 & 12.5             & 21.0                  & 15.7        & 20.6            & -           & -               & -           & -               \\
Additive 2 & 11.9             & 20.2                  & 15.2        & 18.5            & -           & -               & -           & -               \\
Additive 3 & 12.4             & 20.4                  & -           & -               & 15.2        & 18.0            & -           & -               \\
Additive 4 & 12.2             & 20.4                  & 15.3        & 19.3            & -           & -               & -           & -               \\
Additive 5 & 13.5             & 22.7                  & 17.6        & 19.7            & 17.5        & 20.3            & -           & -               \\
Additive 6 & 10.3             & 20.5                  & 14.8        & 20.9            & 15.6        & 19.0            & -           & -               \\
Additive 7 & 12.2             & 20.5                  & 15.4        & 20.7            & 16.1        & 20.4            & 21.2        & 25.1           \\ \hline
\end{tabular}
    \vspace*{1mm}
\label{tab:dist_results}
\end{table}

A key limitation of our method is the model fine-tuned on our ground truth dataset seems unable to distinguish image diversity when image quality is low. Figure \ref{fig:pca_results} indicates Dataset C is the least diverse, while Table \ref{tab:dist_results} indicates the diversity in Dataset C's Additive 7 class is much higher than in any other dataset (seen by the greater distance in the \textit{same} column). Additionally, each of these metrics lacks the ability to distinguish physically accurate features when scoring. Dataset C, though of low image quality, exhibits diverse sets of cell structures that still provide valuable information in the materials discovery platforms (visualized in Appendix \ref{appendix:image_analysis}). 

\section{Conclusion and Future Works}

\label{section:conclusion}
We have identified and evaluated several limitations of current state of the art methods in evaluating generative models from non-traditional problem domains. In particular, we highlight the limitations from using small dataset sizes, images with features different from those in ImageNet, and the need to score based on physically accurate micro-structure formations that can be synthesized in the lab. 

As we progress this work, we hope to leverage additional deep learning tools (e.g. Dragonfly \cite{dragonfly, dragonfly_software}) to statistically quantify image features such as number of cells, cell wall width and size of cells (sample visualization in Appendix \ref{appendix:dragonfly}) and evaluate feasibility of synthesized images. 
Similarly, we plan to leverage known physics-based correlations to embed the knowledge of physical laws that govern our materials dataset during both the learning and evaluation process of the generative models. Further exploring model architectures (e.g. diffusion models), hyper-parameter tuning, and custom loss functions will allow us to improve the quality of synthetic images and provide more accurate evaluation criteria. In general, we hope this work will progress the evaluation of synthetic images in various scientific fields beyond micro-structure imaging where traditional methods may be limited.

\newpage

\begin{ack}
We thank Patrick Blanchard, Rachel Couvreur, Janice Tardiff and Alper Kiziltas for their support on the experimental data generation and discussion. We also thank Ford Motor Company Research \& Advanced Engineering for sponsoring this work. 
\end{ack}


\printbibliography

\newpage

\appendix

\newpage

\section{Datasets Limitations}
\label{appendix:dataset_limitations}

While AI methods require substantially more data than conventional methods, materials scientists have traditionally worked with small \& noisy datasets \cite{ford_paper_currentworkplan}, thanks in part to advancements of generative modeling in small data regimes using methods such as differential augmentation \cite{diff_aug} and adaptive instance normalization \cite{ADAin}. Yet, there is minimal discussion around the impact of such small training datasets on model evaluation metrics. 

The FID score is the Fr\'echet distance $d()$ between the Gaussian with mean $(m, C)$ obtained from $p(.)$ and the Gaussian with mean $(m_w, C_w)$ obtained from $p_w()$ given by the following \cite{fid_score_created}:

\begin{equation}
d^2((m,C), (m_w,C_w)) = {{||m-m_w||}^2_2} + Tr(C + C_w - 2(CC_w)^{1/2})
    \label{equation:fid}
\end{equation}

From Equation \ref{equation:fid}, we see the FID score compares both the mean and variance of the training dataset to the respective components of the synthetic images. The FID score is minimized (ideal) when these variables are as similar as possible. Although any number of synthetic images can be generated, the size of the training dataset can become a limitation. Smaller training datasets provide a smaller variance, thus resulting in a lower FID score when our synthetic images have a similar variance, subsequently encouraging over-fitting.

We ran several trials to highlight this difference, as seen in Table \ref{apdx_tab:fidscores}. The experiments were run using three datasets: CIFAR10, FFHQ* (the thumbnail version), and our internal SEM micro-structure dataset. For each trial, the dataset was randomly (without repetition) split into two groups of varying sizes. These groups were representative of a ground truth (Dataset 1) and a synthetic (Dataset 2) dataset. Note, for the micro-structure trials, Dataset 1 came from our actual ground truth dataset while Dataset 2 was randomly produced using the trained network for Dataset A. As a result, we only see trail results where the ground truth dataset is less than or equal to 1000 images. 

Table \ref{apdx_tab:fidscores} clearly highlights the impact of small training datasets (Dataset 1), even when both datasets are sampled from the same distribution (CIFAR10, FFHQ*). In particular, we notice that the FID score is consistently low, no matter the problem domain, when working with large training and synthetic datasets. However, as the dataset sizes decrease, not only does the FID score diverge quickly, there is a lack of consistency between datasets. As a result, it becomes difficult to use the FID score as a standard metric to compare multiple models during training when working in problem domains with limited data. An alternative metric is needed for scoring generative models. 

\begin{table}[htb]
\centering
\caption{The FID score is reliant on using larger dataset sizes in order to consistently minimize the FID score. As the dataset sizes get smaller, the FID score diverges at different rates depending on the particular dataset. Trial E most closely represents the size of datasets we are working with in our problem domain.
\vspace*{1mm}}
\begin{tabular}{c|cc|ccc}
\hline
\multicolumn{1}{l|}{} & \multicolumn{2}{c|}{Dataset Size} & \multicolumn{3}{c}{FID Score}              \\
Trial                & Dataset 1       & Dataset 2      & CIFAR10     & FFHQ*       & Our SEM Images \\ \hline
A                    & 30,000          & 30,000         & 1.8         & 1.9         & -              \\
B                    & 10,000          & 50,000         & 3.3         & 3.4         & -              \\
C                    & 10,000          & 10,000         & 5.4         & 5.5         & -              \\
D                    & 1000            & 50,000         & 33          & 26          &        62        \\
\textbf{E}           & \textbf{1000}   & \textbf{1000}  & \textbf{54} & \textbf{35} & \textbf{64}      \\
F                    & 100             & 50,000         & 153         & 94          &        76        \\
G                    & 100             & 100            & 180         & 109         &     83          \\ \hline
\end{tabular}
\label{apdx_tab:fidscores}
\end{table}

\section{Model Architecture/Hyper-parameters}
\label{appendix:model_arch}
Dataset A and B were both derived from a base StyleGAN2 model \cite{stylegan2_implementation} with Differential Augmentation \cite{diff_aug}. The augmentations applied were color, cutout, and translation. Dataset A was trained for 280 Kimg and Dataset B for 240 Kimg. Both models were randomly initialized with weights before training. Dataset C is derived from a ResNet18 based Variational Autoencoder.

\newpage
\section{Image Analysis}
\label{appendix:image_analysis}

\begin{figure} [!htb]
\centering
\begin{tabular}{m{1em}ccc}
\rotatebox[origin=l]{90}{Ground Truth Dataset} &
\makecell{\includegraphics[width = 0.3\linewidth]{figures/dataset_base/05pGNP_106.png}} &
\makecell{\includegraphics[width = 0.3\linewidth]{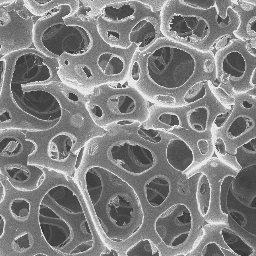}} &
\makecell{\includegraphics[width = 0.3\linewidth]{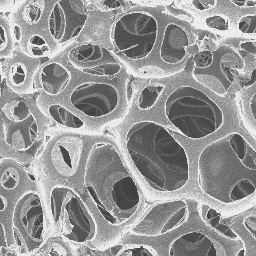}} \\ 

\rotatebox[origin=l]{90}{Dataset A} &
\makecell{\includegraphics[width = 0.3\linewidth]{figures/dataset1/seed0001.png}} &
\makecell{\includegraphics[width = 0.3\linewidth]{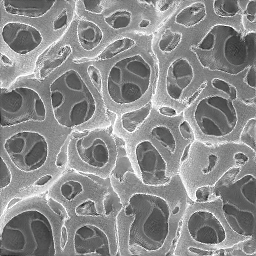}} &
\makecell{\includegraphics[width = 0.3\linewidth]{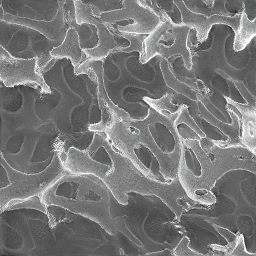}} \\ 

\rotatebox[origin=l]{90}{Dataset B} &
\makecell{\includegraphics[width = 0.3\linewidth]{figures/dataset2/seed0001.png}} &
\makecell{\includegraphics[width = 0.3\linewidth]{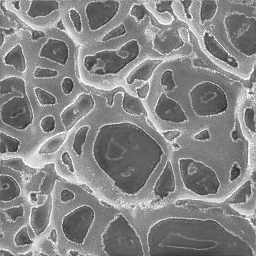}} &
\makecell{\includegraphics[width = 0.3\linewidth]{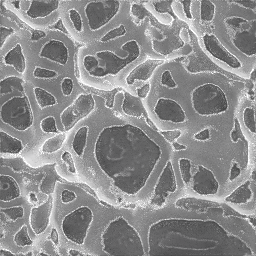}} \\ 

\rotatebox[origin=l]{90}{Dataset C} &
\makecell{\includegraphics[width = 0.3\linewidth]{figures/dataset3/73.png}} &
\makecell{\includegraphics[width = 0.3\linewidth]{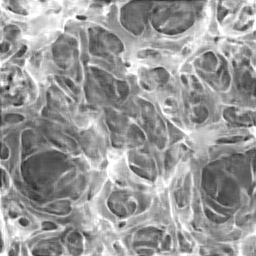}} &
\makecell{\includegraphics[width = 0.3\linewidth]{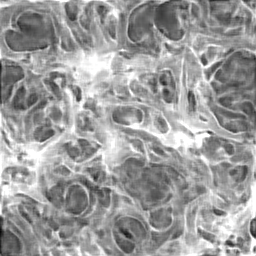}} \\
\end{tabular}
    \vspace*{1mm}
\caption{Specific image analysis on each of the 4 datasets we are working with. Note the variation in detail from each of the datasets. For a viewer outside the material science domain, both Dataset A and Dataset B seem to show very realistic images relative to the Ground Truth Dataset. On the other hand, it is clear the images in Dataset C are not of high quality and are synthetically generated.}
\label{apdx:fig:image_quality}
\end{figure}

\begin{figure} [!htb]
\centering
\begin{tabular}{m{1em}c}
\rotatebox[origin=l]{90}{Ground Truth Dataset} &
\makecell{\includegraphics[width = 0.9\linewidth]{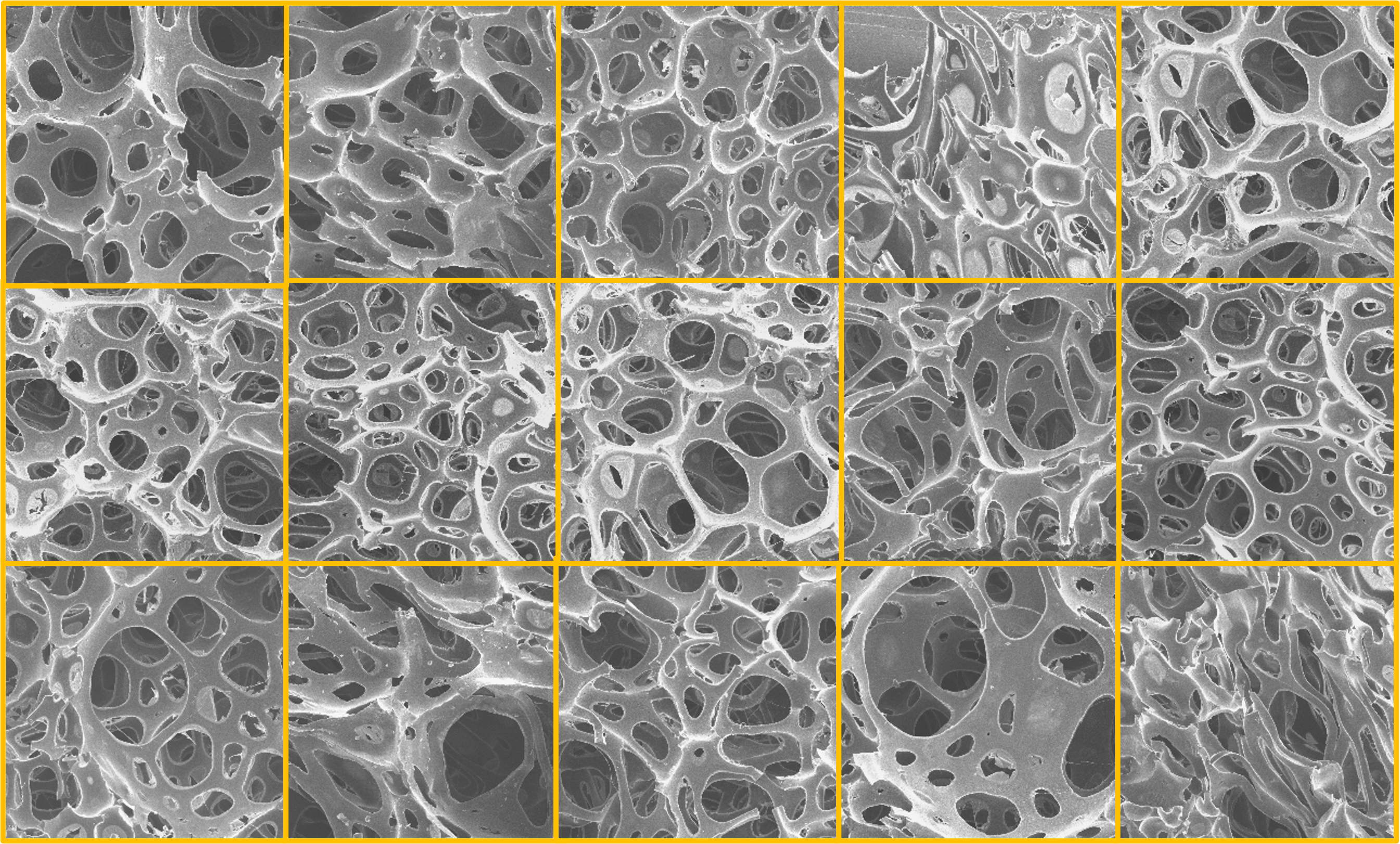}} \\ \\
\rotatebox[origin=l]{90}{Dataset A} &
\makecell{\includegraphics[width = 0.9\linewidth]{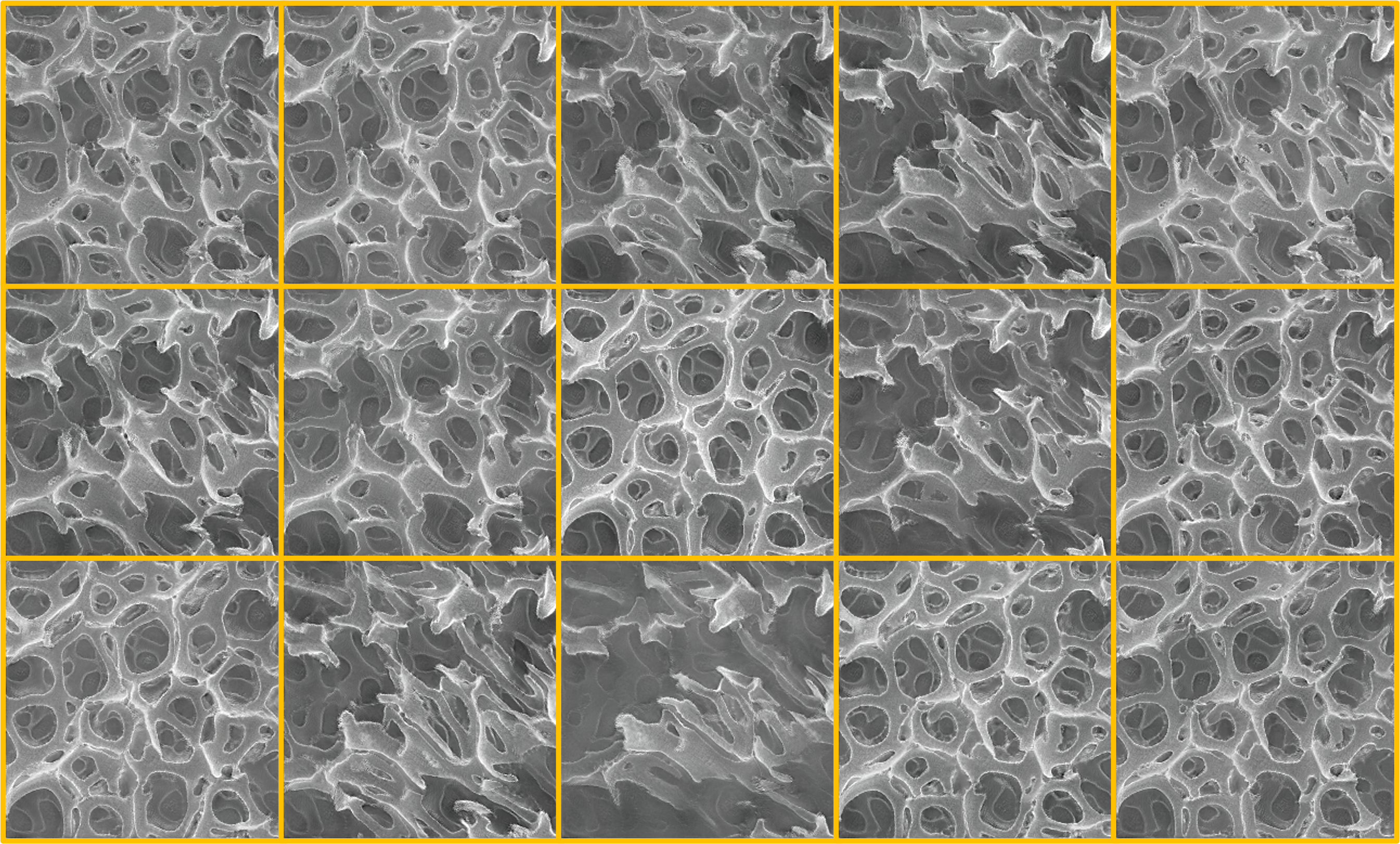}} \\ \\
\end{tabular}
    \vspace*{1mm}
\caption{A random set 15 images were selected from each of the datasets and shown here side by side to highlight image diversity. As expected, the Ground Truth Dataset is extremely diverse as it comes from 7 different combinations of additives. Even repeated SEM images of the same experiment from different angles show different features in the Ground Truth Dataset. On the other hand, we see the model used for generating Dataset B is over fit and repeating the same image over with very minor changes, representative of mode collapse in the generative model. Dataset A provides greater diversity; however, we still notice multiple transitions between the same few phases. Contrastingly, Dataset C is the most diverse with unique micro-structure configurations in each of the images. Though the lower fidelity of images in Dataset C can make it difficult to analyze, we see each image exhibits both unique cell sizes and varying locations for cell connections.}
\end{figure}%
\begin{figure}[!htb] \ContinuedFloat
\centering
\begin{tabular}{cc}
\rotatebox[origin=c]{90}{Dataset B} &
\makecell{\includegraphics[width = 0.9\linewidth]{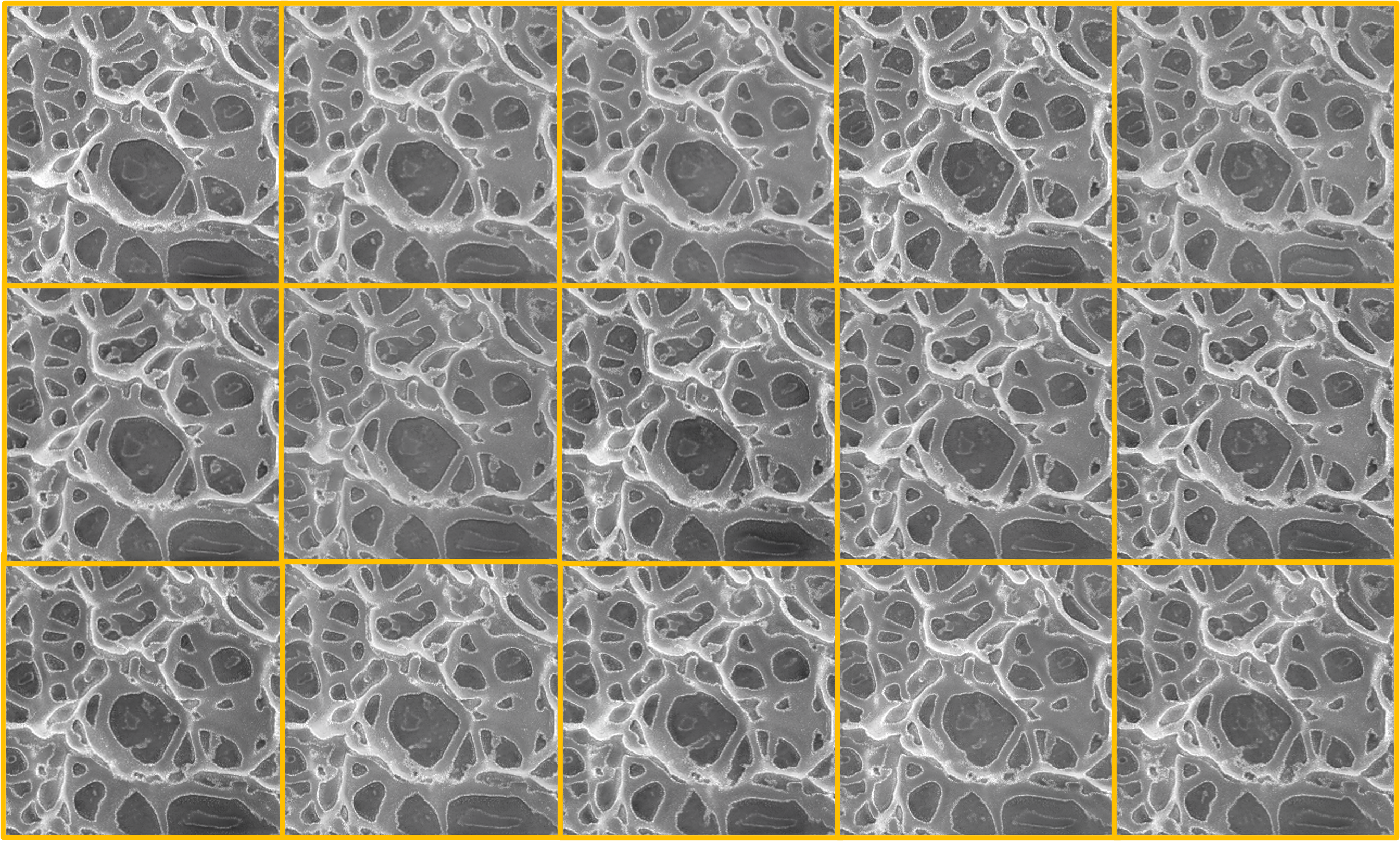}} \\ \\
\rotatebox[origin=c]{90}{Dataset C} &
\makecell{\includegraphics[width = 0.9\linewidth]{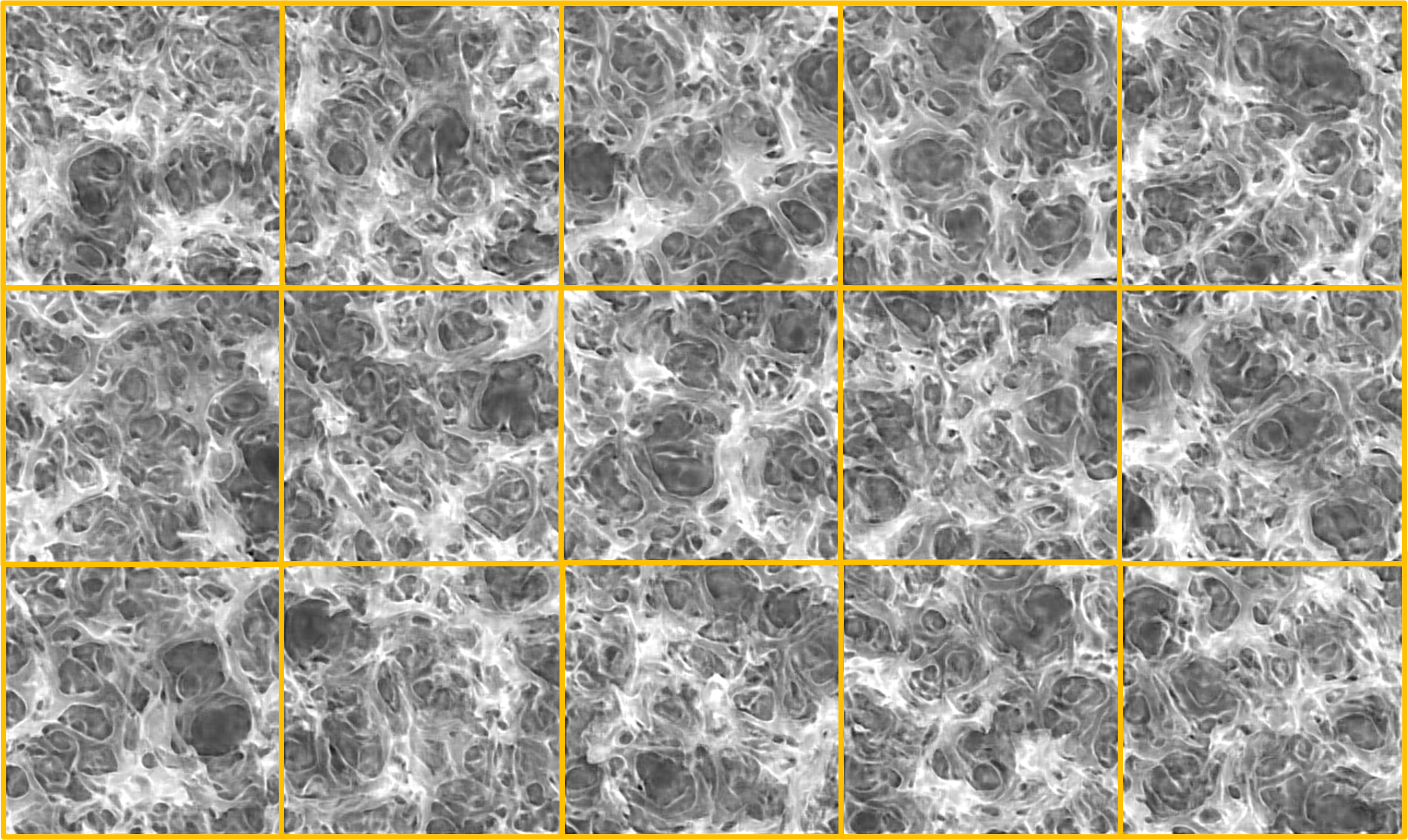}} \\
\end{tabular}
    \vspace*{1mm}
\caption{A random set 15 images were selected from each of the datasets and shown here side by side to highlight image diversity. As expected, the Ground Truth Dataset is extremely diverse as it comes from 7 different combinations of additives. Even repeated SEM images of the same experiment from different angles show different features in the Ground Truth Dataset. On the other hand, we see the model used for generating Dataset B is over fit and repeating the same image over with very minor changes, representative of mode collapse in the generative model. Dataset A provides greater diversity; however, we still notice multiple transitions between the same few phases. Contrastingly, Dataset C is the most diverse with unique micro-structure configurations in each of the images. Though the lower fidelity of images in Dataset C can make it difficult to analyze, we see each image exhibits both unique cell sizes and varying locations for cell connections.}
\label{apdx:fig:image_diversity}
\end{figure}

\begin{figure}[!htb]
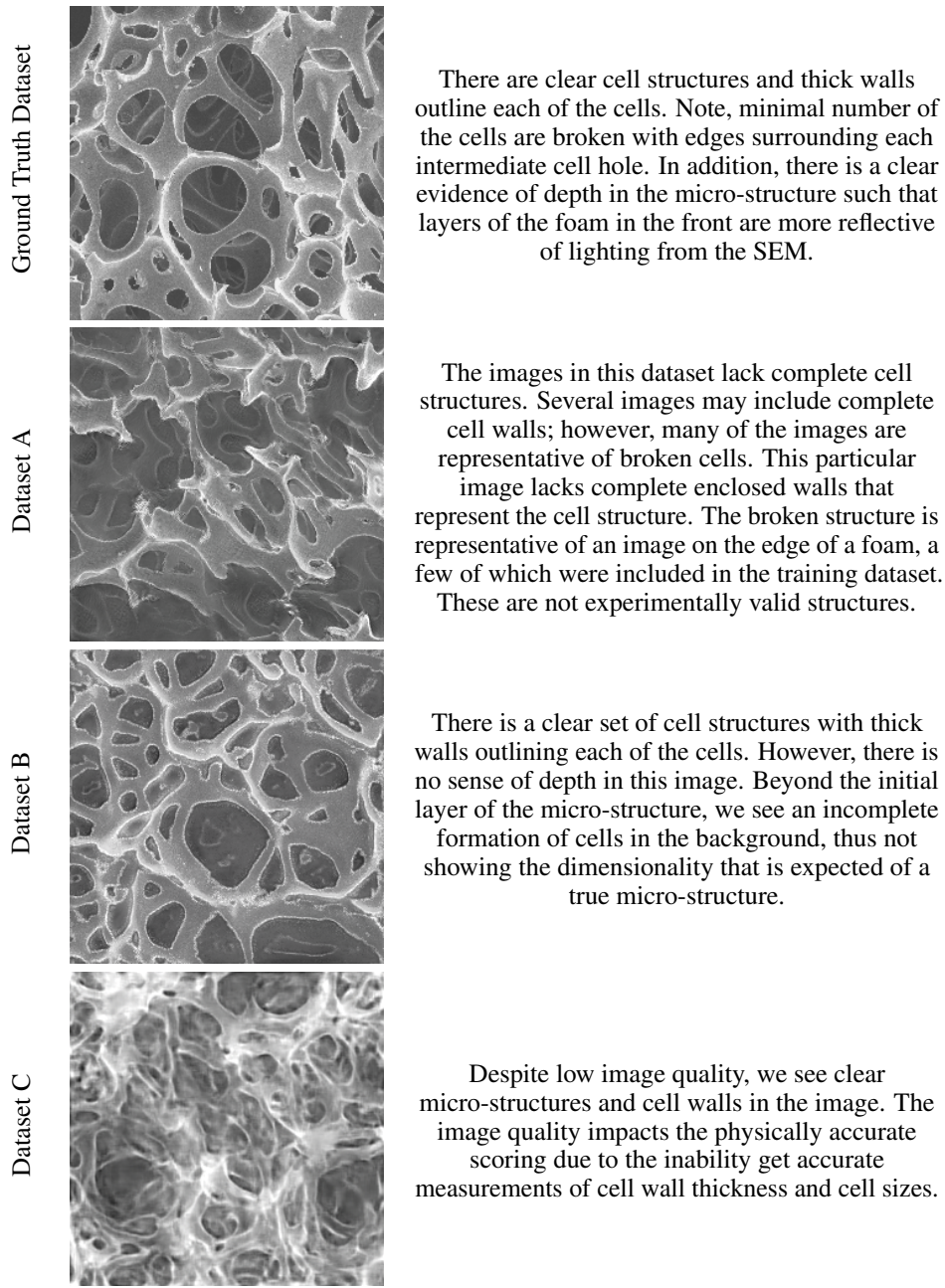

\centering
\begin{tabular}{ m{1em}c
>{\centering\arraybackslash}m{7cm}}
\rotatebox[origin=l]{90}{Ground Truth Dataset} & 
\makecell{\includegraphics[width = 0.3\linewidth]{figures/dataset_base/05pGNP_106.png}} 
& There are clear cell structures and thick walls outline each of the cells. Note, minimal number of the cells are broken with edges surrounding each intermediate cell hole. In addition, there is a clear evidence of depth in the micro-structure such that layers of the foam in the front are more reflective of lighting from the SEM. \\
\rotatebox[origin=l]{90}{Dataset A}            & 
\makecell{\includegraphics[width = 0.3\linewidth]{figures/dataset1/seed0003.png}}
& The images in this dataset lack complete cell structures. Several images may include complete cell walls; however, many of the images are representative of broken cells. This particular image lacks complete enclosed walls that represent the cell structure. The broken structure is representative of an image on the edge of a foam, a few of which were included in the training dataset. These are not experimentally valid structures.                        \\
\rotatebox[origin=l]{90}{Dataset B}          & 
\makecell{\includegraphics[width = 0.3\linewidth]{figures/dataset2/seed0002.png}}
& There is a clear set of cell structures with thick walls outlining each of the cells. However, there is no sense of depth in this image. Beyond the initial layer of the micro-structure, we see an incomplete formation of cells in the background, thus not showing the dimensionality that is expected of a true micro-structure.                    \\
\rotatebox[origin=l]{90}{Dataset C}           &  
\makecell{\includegraphics[width = 0.3\linewidth]{figures/dataset3/74.png}}
& Despite low image quality, we see clear micro-structures and cell walls in the image. The image quality impacts the physically accurate scoring due to the inability get accurate measurements of cell wall thickness and cell sizes.
\end{tabular}
\caption{A descriptive analysis on the physically accurate scoring of each of the datasets. The key considerations which are visible to the human eye in an SEM image include cell structures, cell wall, and depth of cells beyond initial layer. We only consider visually interpretable conclusions currently; however, there is opportunity to leverage scoring metrics, as seen in Appendix \ref{appendix:dragonfly}, to quantify such measurements.}
\label{apdx:fig:image_physically_acc}
\end{figure}

\clearpage

\section{Dragonfly Image Segmentation}
\label{appendix:dragonfly}

\begin{figure}[!h]
    \centering
    \subfloat[\centering Manual segmentation using Dragonfly Deep Learning Tool]{
        {\includegraphics[width=.4\linewidth]{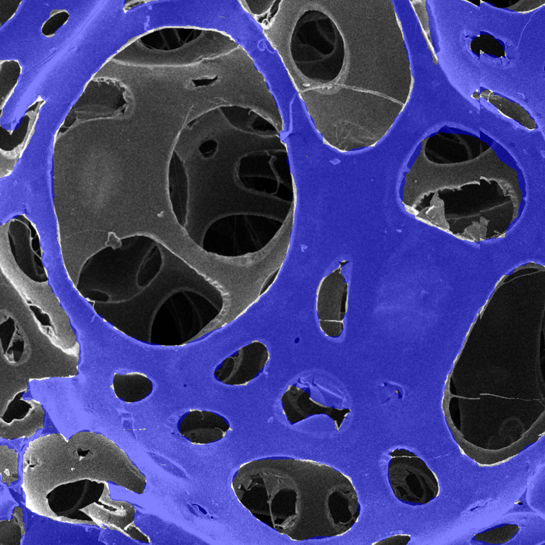} }
    }
    \enspace
    \subfloat[\centering Otsu thresholding of internal cell structure]{
        {\includegraphics[width=.4\linewidth]{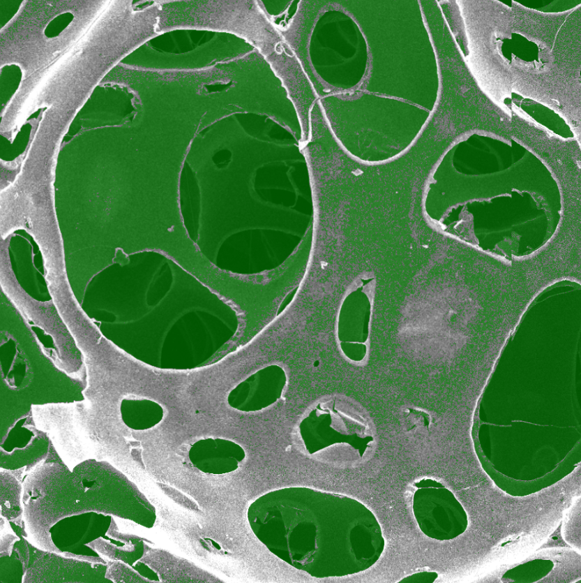} }
    }
    \\
    \subfloat[\centering Segmentation map prediction of pores and cell structure]{
        {\includegraphics[width=.4\linewidth]{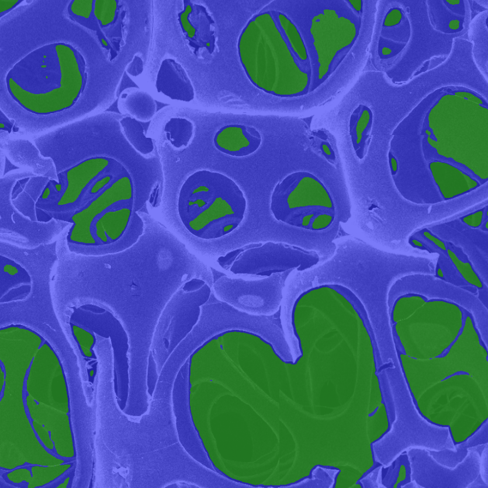} }
    }
        \vspace*{1mm}
    \caption{Dragonfly, a deep learning based image processing software, allows us to perform automated labeling on micro-structure images with high precision given very limited data. Beyond image segmentation, this tool allows us to quantify parameters such as number of cells, cell wall width and size of cells. Using such values, we can statistically score various additive types as well as synthetic images features.}
    \label{apdx:fig:dragonfly_results}
    \vspace*{-2mm}
\end{figure}

\end{document}